# SKY-MAPS OF THE SIDEREAL ANISOTROPY OF GALACTIC COSMIC RAY INTENSITY AND ITS ENERGY DEPENDENCE

K. MUNAKATA[1], N. MATSUMOTO[1], S. YASUE[1], C. KATO[1], S. MORI[1],
M. TAKITA[2], M. L. DULDIG[3], J. E. HUMBLE[4] AND J. KÓTA[5]

[1]*Physics Department, Shinshu University, Matsumoto, Nagano 390-8621, Japan*
[2] *Institute for Cosmic Ray Research, University of Tokyo, Kashiwa, Japan*
[3]*Australian Antarctic Division, Kingston, Tasmania 7050, Australia*
[4]*School of Mathematics and Physics, University of Tasmania, Hobart, Tasmania 7001, Australia*
[5]*Lunar and Planetary Laboratory, University of Arizona, Tucson, Arizona, USA*

We analyze the sidereal daily variations observed between 1985 and 2006 at Matsushiro, Japan (MAT) and between 1993 and 2005 at Liapootah, Tasmania (LPT). These stations comprise the two hemisphere network (THN) of underground muon detectors in Japan and Australia. Yearly mean harmonic vectors at MAT and LPT are more or less stable without any significant change in phase and amplitude in correlation with either the solar activity- or magnetic-cycles. In this paper, therefore, we analyze the average anisotropy over the entire observation periods, i.e. 1985-2006 for MAT and 1993-2005 for LPT. We apply to the THN data a best-fitting analysis based on a model anisotropy in space identical to that adopted by Amenomori *et al.* (2007) for Tibet III data. The median energies of primary cosmic rays recorded are ~0.5 TeV for THN and ~5 TeV for the Tibet III experiment. It is shown that the intensity distribution of the best-fit anisotropy is quite similar to that derived from Tibet III data, regardless of the order of magnitude difference in energies of primary particles. This, together with the THN observations, confirms that the analysis by Amenomori *et al.* (2007) based on the Tibet III experiment in the northern hemisphere is not seriously biased. The best-fit amplitudes of the anisotropy, on the other hand, are only one third or less of those reported by the Tibet III experiment, indicating attenuation due to solar modulation. The rigidity dependence of the anisotropy amplitude in the sub-TeV region is consistent with the spectrum reported by Hall et al. (1999), smoothly extending to the Tibet III result in the multi-TeV region. The amplitude at higher energies appears almost constant or gradually decreasing with increasing rigidity. The rigidity spectrum indicating the solar modulation also supports the conclusion first implied by the Super Kamiokande deep underground experiment that the large scale anisotropy observed by Tibet III is due to the charged component of primary cosmic rays, and not due to high energy gamma rays to which underground muon detectors have negligible response.

## 1. Introduction

Long term observations of the count rates of cosmic rays recorded by underground muon telescopes have consistently reported the existence of an average diurnal variation in sidereal time [1-5]. Historically, this variation has been studied in an attempt to elucidate the origin of galactic cosmic rays (GCRs) with energies between a few hundred and a few thousand GeV. The observations from underground muon telescopes also provide interesting comparisons with those made from air-shower experiments, which respond to primary cosmic rays with energies more than an order of magnitude higher [6-7]. Most investigations have concluded that the variation is small (~0.1%) and the apparent right ascension of intensity maximum is somewhere in the early hours of the local sidereal day. Presently, no common consensus exists about the production mechanism of this variation in the local region of the galaxy. Information on the declination of the anisotropy can only be obtained by making multiple observations of various regions of the sky. This was not possible until recently. With multi-directional telescopes, the amplitude of this variation was shown to be distributed asymmetrically about the celestial equator [3]. It has now been conclusively demonstrated that the sidereal daily variation will be recorded with the largest amplitude in its first (diurnal) harmonic when viewed from around mid-latitudes in the southern hemisphere [8-9]. This is called the North-South (NS) asymmetry of the sidereal diurnal variation.

Aiming to reveal detail features of this NS asymmetry for understanding the physical mechanism responsible for such an asymmetry, the two-hemisphere network (THN) of underground muon telescopes at Matsushiro (at a vertical depth of 220 meters of water equivalent: m.w.e.) in central Japan and at Liapootah (154 m.w.e.) in Tasmania, Australia started operation in December, 1991 and continuously monitored the high energy GCR intensity until observations at Liapootah ended in March 2006. By using the initial four years of data from the THN, together with other underground muon data available, Hall *et al.* [10-11] used non-linear iterative fitting procedures to model the observed daily variations by a superposition of two Gaussian functions. Based on the best-fit parameters, they plotted the observed intensity distributions in a contour map on the celestial sphere and found some notable features, particularly in the intensity excess region. First, this region had its maximum shifted significantly southward over the equator and, second, its distribution is not symmetric around the maximum direction, appearing "squashed" along a plane which is not aligned with either the solar or galactic equatorial planes. The physical implications of these features are still unknown.

By analyzing GCRs recorded by the Tibet III air shower experiment, Amenomori *et al*. [12] first presented a precise sky-map of GCR intensity in the multi-TeV region. In the sky-map representing the relative intensity as a function of the declination and right ascension of the incident direction, the large-scale structure of the anisotropy is evident with a statistical significance exceeding 10 sigma. Amenomori *et al*. [13] (hereafter referred to as paper I) proposed a model interpreting this large-scale anisotropy in terms of the local interstellar structure around the heliosphere. The heliosphere is located inside the local interstellar cloud (LIC) very close to the inner edge of the LIC [14]. If the GCR population is lower inside the LIC than outside, uni-directional and bi-directional flows (UDF and BDF) are both expected from the parallel diffusion of GCRs into the LIC along the local interstellar magnetic field (LISMF) connecting the heliosphere with the region outside the LIC, where the GCR population is higher. In addition, the UDF perpendicular to the LISMF is also expected from the $\mathbf{B} \times \nabla n$ drift anisotropy driven by a spatial gradient of GCR density ($n$) in the LISMF ($\mathbf{B}$). If the weak scattering regime applies, the contribution from the cross-field diffusion to the perpendicular UDF should be much smaller than the drift anisotropy. It is found that the LISMF orientation deduced from the best-fit direction of the BDF axis is almost parallel to the galactic plane and is more consistent with the suggestion of Frisch [15] than with that of Lallement *et al.* [16].

The Tibet III experiment is currently the world's highest precision measurement of GCR intensity in this energy region, utilizing both high count rates and good angular resolution of the incident direction. The GCR anisotropy in this energy region is free from solar modulation, while it is still sensitive to the local magnetic field structure with a spatial scale comparable to the Larmor radius of GCR particles, about 0.002 pc or 300 AU for 5 TeV protons in the 3 μG LISMF. The analysis of Tibet III data by paper I, on the other hand, can in principle be biased, because the experiment cannot observe the mid-latitude sky in the southern hemisphere, where a large anisotropy is reported as a consequence of the NS asymmetry of the sidereal diurnal variation [10-11]. In order to remove such a bias, it is necessary to analyze the existing data in the southern hemisphere together with the Tibet AS data.

In this paper, we analyze the THN data collected during its total observation period from 1993 to 2005. We apply to the THN data a best-fitting analysis identical to that adopted by paper I for the Tibet III data and derive the energy dependence of the anisotropy in space by comparing the best-fit parameters from the THN observations in the sub-TeV region and the Tibet III experiment in the multi-TeV region.

## 2. Analysis

### 2.1 Long term variation of the sidereal anisotropy

We analyze the sidereal daily variations observed with a pair of muon detectors at Matsushiro (MAT) between 1985 and 2006 in the northern hemisphere and at Liapootah (LPT) between 1993 and 2005 in the southern hemisphere. These two detectors are both multidirectional and capable for measuring muon intensities in a total of 34 directional channels simultaneously. The median primary energy for muons recorded by these detectors covers the sub-TeV region ranging from 0.565 to 0.861 TeV in MAT and from 0.454 to 0.984 TeV in LPT, respectively, while the effective viewing latitude corrected for geomagnetic bending ranges from 61.3°N to 15.0°S in MAT and from 13.6°N to 57.6°S in LPT. For detailed information of these detectors, readers can refer to Hall *et al.* [10-11]. We first obtain the first and second harmonics, $A_{1\,i,j}$, $B_{1\,i,j}$, $A_{2\,i,j}$ and $B_{2\,i,j}$, by expanding the yearly mean sidereal daily variation $D_{i,j}(t)$ in the $j$-th directional channel ($j$ =1 to 34) of the $i$-th detector ($i$ =1 for MAT and $i$ =2 for LPT), as

$$D_{i,j}(t) = \sum_{m=1}^{2}\left(A_{m\,i,j}\cos m\omega t_i + B_{m\,i,j}\sin m\omega t_i\right), \qquad (1)$$

where $t_i$ is the local sidereal time for the $i$-th detector and $\omega = 2\pi/24$. Figure 1 displays $A_{m\,i,j}$ and $B_{m\,i,j}$ in five directional channels viewing respectively the vertical (V), north (N), south (S), east (E) and west (W) directions as harmonic vectors. The amplitude of each vector is represented by its length from its origin, while the phase is represented by the angle measured clockwise from the vertical ($y$) axis, i.e. the local times of $+y$, $+x$, $-y$, and $-x$ directions in Figures 1a-1b are, respectively, 00:00, 06:00, 12:00, and 18:00. To demonstrate the long-term variation, we plot in this figure "vector summation dials", in which yearly mean vectors are summed one by one from the first year, i.e. 1985 for MAT and 1993 for LPT. To show the difference in the average amplitudes by MAT and LPT free from the difference in observation periods, we scaled each annual vector by dividing it by the number of years of observation (22 for MAT and 13 for LPT). The final data point in each summation dial, therefore, represents the harmonic vector averaged over the total observation period. The errors attached to the vertical vector (V) are deduced from the dispersion of the yearly values of $A_{m\ i,j}$ and $B_{m\ i,j}$. It is seen first in this figure that the average harmonic vectors by MAT and LPT are both statistically significant and persistent resulting in the summation dial extending toward almost the same local time. It is also seen, in common for both detectors that the phase of the vector in the east viewing channel (E) is systematically earlier than that in the west viewing channel (W) with the phase in the vertical channel (V) being intermediate. This is consistent with the daily variation due to the anisotropy recorded in directional channels on the spinning Earth. On the other hand, there are also notable features appearing differently in MAT in the northern hemisphere and LPT in the southern hemisphere. The average vector in the south viewing channel (S) is much larger than that in the north viewing channel (N) in MAT. There is actually a geometrical effect so that one does expect larger anisotropy for MAT-S than for MAT-N, but the difference is larger and cannot be explained by the effect. Furthermore, all five vectors (V, N, S, E, W) have almost the same amplitude in LPT, different to those vectors in MAT. This is the signature of the north-south asymmetry of the sidereal anisotropy as observed with multidirectional detectors [8-9].

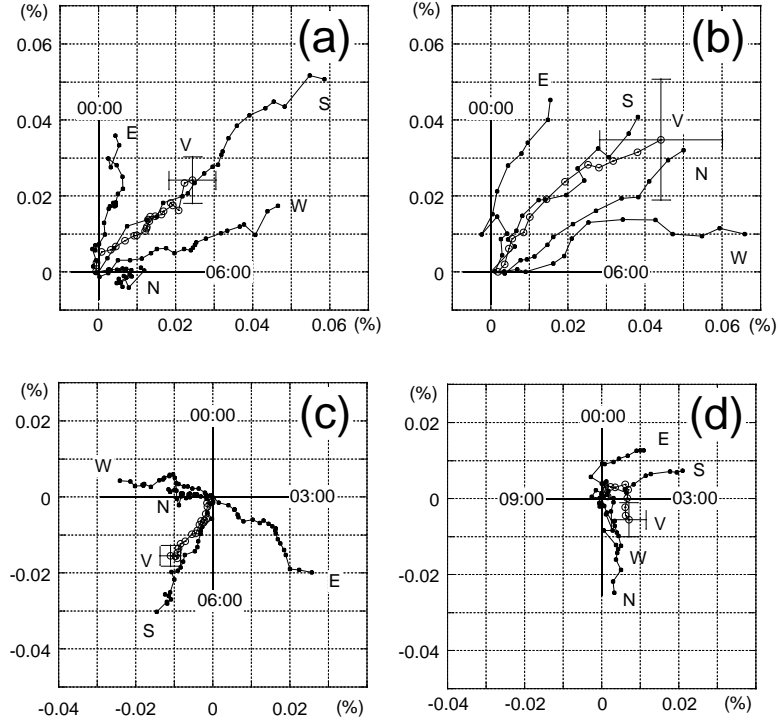

Fig. 1. Summation dials of the yearly mean harmonic vectors observed by five directional channels of MAT and LPT. Directional channels are indicated by alphabetical characters, "V", "N", "S", "E", "W". The upper panels (a), (b) display the diurnal vectors ($m = 1$), while the lower panels (c), (d) show the semi-diurnal vectors ($m = 2$). The left panels (a), (c) show observations by MAT during the 22 years between 1985 and 2006, while the right panels (b), (d) show observations by LPT during the 13 years between 1993 and 2005. To allow direct comparison of the average vectors for MAT and LPT, free from the difference in observation periods, each annual vector is scaled by the number of years of observation for that station. The error attached to each vector labeled "V" represents the standard deviation of the long-term mean value, calculated from the dispersion of the annual vectors.

As shown in Figure 1, for all the directional channels at both MAT and LPT there are few, if any, significant changes in amplitude and phase from year to year. In this paper, therefore, we analyze the average anisotropy over the whole observation period, i.e. 1985-2006 for MAT and 1993-2005 for LPT. We will analyze the temporal variation of the anisotropy possibly due to solar modulation correlated with solar activity- and magnetic-cycles in a separate paper.

## 2.2 Best-fitting with the model anisotropy

Following paper I, we assume the GCR intensity in free space consisting of the bi-directional and uni-directional flows (BDF and UDF) along a reference axis in the direction of the right ascension $\phi_R$ and co-latitude $\theta_R$ in celestial coordinates. We also assume another UDF along the direction $(\theta_R^\perp, \phi_R^\perp)$ perpendicular to the axis. In this model, the anisotropic GCR intensity $F(\theta_J, \phi_J; \theta_R, \phi_R, \theta_R^\perp, \phi_R^\perp)$ expected in a directional channel viewing right ascension $\phi_J$ and co-latitude $\theta_J$ is given as

$$F(\theta_J, \phi_J; \theta_R, \phi_R, \theta_R^\perp, \phi_R^\perp) = \eta_1^\perp P_1^0(\cos\chi^\perp) + \eta_1^\parallel P_1^0(\cos\chi) + \eta_2^\parallel P_2^0(\cos\chi), \quad (1)$$

where $\eta_2^\parallel$ is the amplitude of the BDF, $\eta_1^\parallel$ and $\eta_1^\perp$ are respectively amplitudes of UDFs parallel and perpendicular to the BDF axis and $\chi$ and $\chi^\perp$ are angles of the viewing direction measured from the BDF axis and the perpendicular UDF axis, respectively. $P_1^0$ and $P_2^0$ in Eq. (1) are the Schmidt semi-normalized associated Legendre functions, defined as

$$P_n^m(\cos\theta) = \begin{cases} P_{n,m}(\cos\theta) & \text{for } m = 0 \\ \sqrt{\frac{2(n-m)!}{(n+m)!}} P_{n,m}(\cos\theta) & \text{for } m \neq 0 \end{cases} \quad (2)$$

and

$$P_{n,m}(\cos\theta) = \sin^m\theta \frac{d^m P_n(\cos\theta)}{d(\cos\theta)^m}. \quad (3)$$

Angles $\chi$ and $\chi^\perp$ in Eq. (1) are expressed in terms of the right ascensions ($\phi_R$ and $\phi_R^\perp$) and co-latitudes ($\theta_R$ and $\theta_R^\perp$) of the BDF and the perpendicular UDF axes, as

$$\cos\chi = \cos\theta_R\cos\theta_J + \sin\theta_R\sin\theta_J\cos(\phi_J - \phi_R), \quad (4)$$
$$\cos\chi^\perp = \cos\theta_R^\perp\cos\theta_J + \sin\theta_R^\perp\sin\theta_J\cos(\phi_J - \phi_R^\perp). \quad (5)$$

By introducing Eqs. (4) and (5) into (1) and eliminating terms which are independent of $\phi_J$, we get the anisotropy responsible for the daily variation, as

$$\begin{aligned} f(\theta_J, \phi_J; \theta_R, \phi_R, \theta_R^\perp, \phi_R^\perp) = \\ (x_1^1\cos\phi_J + y_1^1\sin\phi_J)P_1^1(\cos\theta_J) \\ + (x_2^1\cos\phi_J + y_2^1\sin\phi_J)P_2^1(\cos\theta_J) \\ + (x_2^2\cos 2\phi_J + y_2^2\sin 2\phi_J)P_2^2(\cos\theta_J), \end{aligned} \quad (6)$$

where $x_1^1$, $y_1^1$, $x_2^1$, $y_2^1$, $x_2^2$ and $y_2^2$ are called space harmonic components and given, as

$$x_1^1 = \eta_1^\perp P_1^1(\cos\theta_R^\perp)\cos\phi_R^\perp + \eta_1^\parallel P_1^1(\cos\theta_R)\cos\phi_R, \quad (7a)$$
$$y_1^1 = \eta_1^\perp P_1^1(\cos\theta_R^\perp)\sin\phi_R^\perp + \eta_1^\parallel P_1^1(\cos\theta_R)\sin\phi_R, \quad (7b)$$
$$x_2^1 = \eta_2^\parallel P_2^1(\cos\theta_R)\cos\phi_R, \quad (7c)$$
$$y_2^1 = \eta_2^\parallel P_2^1(\cos\theta_R)\sin\phi_R, \quad (7d)$$
$$x_2^2 = \eta_2^\parallel P_2^2(\cos\theta_R)\cos 2\phi_R, \quad (7e)$$
$$y_2^2 = \eta_2^\parallel P_2^2(\cos\theta_R)\sin 2\phi_R. \quad (7f)$$

The daily variation $d_{i,j}(t)$ expected for the $j$-th directional channel of the $i$-th detector from $f(\theta_J, \phi_J; \theta_R, \phi_R, \theta_R^\perp, \phi_R^\perp)$ in Eq. (6) is given, as

$$d_{i,j}(t) = \sum_{m=1}^2 (a_{m\,i,j}\cos m\omega t_i + b_{m\,i,j}\sin m\omega t_i), \quad (8)$$

where $a_{m\,i,j}$ and $b_{m\,i,j}$ are the expected harmonic components, related to the space harmonic components by the coupling coefficients, $c_{ni,j}^m$ and $s_{ni,j}^m$, as

$$a_{1i,j} = c^1_{1i,j} x^1_1 + s^1_{1i,j} y^1_1 + c^1_{2i,j} x^1_2 + s^1_{2i,j} y^1_2, \tag{9a}$$

$$b_{1i,j} = -s^1_{1i,j} x^1_1 + c^1_{1i,j} y^1_1 - s^1_{2i,j} x^1_2 + c^1_{2i,j} y^1_2, \tag{9b}$$

$$a_{2i,j} = c^2_{2i,j} x^2_2 + s^2_{2i,j} y^2_2, \tag{9c}$$

$$b_{2i,j} = -s^2_{2i,j} x^2_2 + c^2_{2i,j} y^2_2. \tag{9d}$$

We calculate the coupling coefficients by utilizing the response function of the atmospheric muons to the primary particles by Murakami *et al.* [17] and assuming $P^{0.7}$ dependence of the anisotropy on the rigidity $P$ of the primary GCR reported from the analysis of the initial THN data [10-11].

By comparing $a_{m\,i,j}$ and $b_{m\,i,j}$ with the observed harmonic components $A_{m\,i,j}$ and $B_{m\,i,j}$, we obtain six best-fit parameters ($\eta^\perp_1$, $\eta^\parallel_1$, $\eta^\parallel_2$, $\theta^\perp_R$, $\theta_R$, $\phi_R$) minimizing the residual $S$ defined, as

$$S = \sum_{i,j} \left\{ \frac{(A_{1i,j} - a_{1i,j})^2}{\sigma_{A_{1i,j}}{}^2} + \frac{(B_{1i,j} - b_{1i,j})^2}{\sigma_{B_{1i,j}}{}^2} + \frac{(A_{2i,j} - a_{2i,j})^2}{\sigma_{A_{2i,j}}{}^2} + \frac{(B_{2i,j} - b_{2i,j})^2}{\sigma_{B_{2i,j}}{}^2} \right\}, \tag{10}$$

where $\sigma_{A_{m\,i,j}}$ and $\sigma_{B_{m\,i,j}}$ are respectively errors of $A_{m\,i,j}$ and $B_{m\,i,j}$ deduced from standard deviations of yearly mean values. Note that, if $\theta^\perp_R$, $\theta_R$, $\phi_R$ are selected, $\phi^\perp_R$ is uniquely given by the orthogonal condition. For a selected set of $\theta^\perp_R$, $\phi^\perp_R$, $\theta_R$, $\phi_R$, $\eta^\perp_1$, $\eta^\parallel_1$, $\eta^\parallel_2$ can be determined by a simple linear least-square method. It is noted that this analysis gives us the best-fit anisotropy in space, being corrected for the geomagnetic and atmospheric effects by utilizing the coupling coefficients in Eqs.(9a)-(9d). This is an important difference from the previous analysis by Hall *et al.* [10-11], who presented the model anisotropy best-fit to the observed daily variation itself tentatively ignoring terrestrial effects.

## 2.3 Best-fit result

The best-fit parameters to the THN data are listed and compared with those by paper I in Table 1. It is seen in this table that the best-fit orientation of the reference axis of the anisotropy is quite similar for the two independent observations at 0.5 TeV by the THN and at 5 TeV by the Tibet III experiment, with the angular difference equal to or less than 10°. This implies, indirectly, that the conclusion derived by paper I from the best-fit analysis to Tibet III data covering only the northern hemisphere is not seriously biased.

Table 1. Best-fit parameters for THN data at the median primary rigidity ($P_m$) of 0.5 TV (upper line). The parameters derived by paper I for Tibet III data at $P_m = 5.0$ TV are also listed for comparison (lower line). Errors of amplitudes are only statistical.

| $P_m$ (TV) | $\eta^\perp_1$ ($\times 10^{-2}$) | $\delta^\perp_R$ (°) | $\phi^\perp_R$ (°) | $\eta^\parallel_1$ ($\times 10^{-2}$) | $\eta^\parallel_2$ ($\times 10^{-2}$) | $\delta_R$ (°) | $\phi_R$ (°) | $\chi^2/$ $d.o.f.$ |
|---|---|---|---|---|---|---|---|---|
| 0.5 | 0.020±0.001 | -12.5 | 2.5 | 0.016±0.001 | 0.020±0.001 | -17.1 | 96.4 | 1.72 |
| 5.0 | 0.108±0.001 | -22.5 | 357.5 | 0.096±0.002 | 0.095±0.001 | -22.5 | 97.4 | 2.44 |

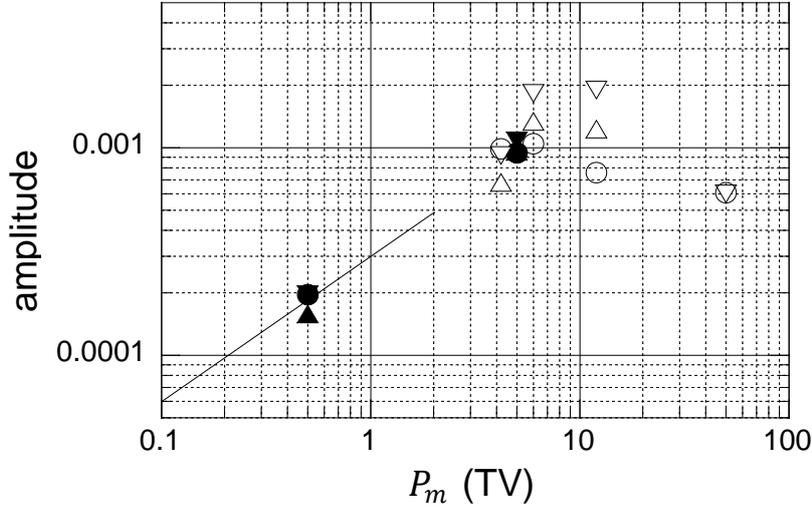

Fig. 2. Rigidity spectra of the best-fit anisotropy. Best-fit amplitudes $\eta_1^{\parallel}$, $\eta_1^{\perp}$ and $\eta_2^{\parallel}$ are respectively displayed by full inverted triangles, full triangles and full circles, each as a function of the median primary rigidity ($P_m$) which is 5.0 TV for Tibet III experiment and normalized to 0.5 TV for THN. Each open symbol displays corresponding amplitude derived from best-fitting to each of five subset of Tibet III data with different median rigidities of 4.2, 6.0, 12.0 and 50.0 TV. The thin solid line indicates the spectrum ($P_m^{0.7}$) reported by Hall *et al.* [11].

The amplitudes of the best-fit anisotropy seen by the THN, on the other hand, are only one third or less of those observed by Tibet III, indicating an attenuation probably due to solar modulation effects. It is seen in Figure 2, showing the amplitude as a function of the median primary rigidity ($P_m$), that the rigidity spectrum ($P_m^{0.7}$) reported by Hall *et al.* [11] and used in this paper is fairly consistent with the Tibet III experiment at 5 TeV. The amplitude at higher rigidities appears almost constant or gradually decreasing above 10 TeV, as shown by open symbols in the figure indicating the best-fit amplitudes to five subset data with different energies.

In Figure 3, we show the best-fit performance of the model anisotropy for sidereal daily variations observed with the THN. Each panel of this figure displays the sidereal daily variation recorded in one of eight directional channels in the THN. It is seen first that the observed daily variation in each directional channel is well reproduced by Eq. (1), i.e. a superposition of the diurnal ($m = 1$) and semi-diurnal ($m = 2$) variations displayed by the dotted curve. The solid curve, which indicates the reproduction by the best-fit model described in the preceding section, also shows reasonable agreement with the data, particularly in directional channels viewing the equatorial region with smaller declination. For directional channels with larger declination, on the other hand, there is notable deviation of the best-fit model from the data. For directional channels viewing high northern (southern) latitudes, the phase of the solid curve deviates to earlier (later) hours from the data and the dotted curve. This is due to the observed intensity maximum around ~06:00 local sidereal time, which has the maximum phase shifting to earlier hours as the viewing latitude moves southward over the equator, as first reported by Hall *et al.* [10-11]. This feature is also reported with a great significance by the Tibet III experiment [18] and cannot be reproduced properly by the simple model anisotropy assumed in this paper (see also paper I).

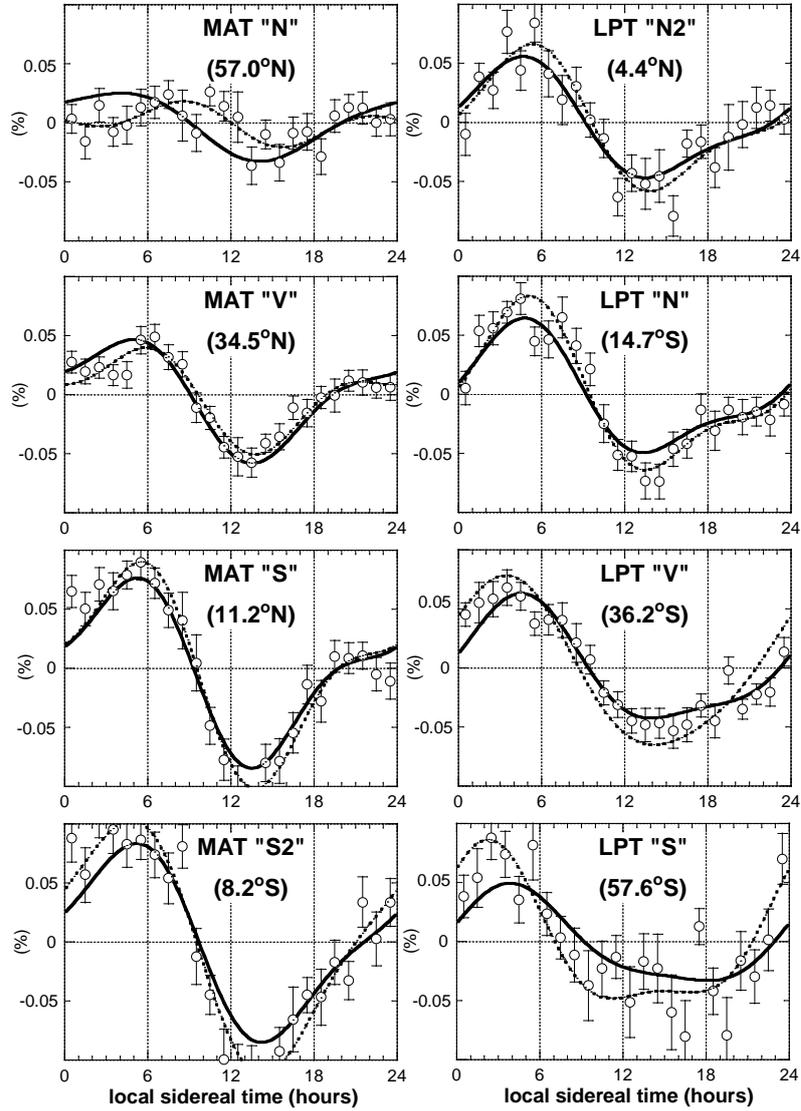

Fig. 3. Best-fit performance of the model anisotropy for sidereal daily variations observed with the THN. Each panel of this figure displays the sidereal daily variation in % recorded in each of eight directional channels of MAT (left) and LPT (right). The name of directional channel and its average declination of viewing are indicated in each panel. Error bars are deduced from the standard deviation of yearly mean count rate at each local sidereal time. The dotted curve displays the superposition of the diurnal ($m = 1$) and semi-diurnal ($m = 2$) variations, while the solid curve displays the reproduction by the best-fit model described in the text.

## 3. Summary and discussion

It is now well established by experiments in the multi-TeV region that the amplitude of the anisotropy is about 0.1 % or $10^{-3}$. Based on the diffusive propagation of GCRs throughout the large-scale galactic magnetic field, it has often been discussed that the amplitude ($\eta_\parallel$) is given as

$$\eta_\parallel \sim 3\kappa_\parallel/(Lc), \tag{11}$$

where $\kappa_\parallel$ is the diffusion coefficient, $L$ is the scale length of the region where the diffusive propagation takes place and $c$ is the speed of light. Using $10^{-3}$ for $\eta_\parallel$ and $10^{29}$ cm$^2$/s for $\kappa_\parallel$ for 1 TeV GCRs [19], we obtain $L \sim 10^{22}$ cm = 3.2 kpc which is comparable to the scale length of the entire Galaxy. A difficulty in interpreting Tibet III's anisotropy in terms of this large-scale diffusion picture is that the observed angular separation between the excess and the deficit is only ~120°, which is much smaller than the 180° expected from a uni-directional flow (UDF) due to simple diffusion. The separation angle, on the other hand, is significantly larger than the 90° expected from a bi-directional flow (BDF) which is expected in the direction along the magnetic field from the adiabatic focusing or loss-cone effect. Only a combination of UDF and BDF can achieve a reasonable fit to the observed anisotropy, as reported by paper I.

As shown in Table 1, the best-fit analysis indicates that there is also a UDF with amplitude ($\eta_\perp$) of ~$10^{-3}$ (at 5 TeV) perpendicular to the BDF, i.e. perpendicular to the magnetic field, if the orientation of BDF is parallel to the field. The UDF perpendicular to the magnetic field is generally produced either from cross-field diffusion or the drift effect. The mean free path ($\lambda_\parallel$) of the pitch-angle scattering responsible for the diffusion is given as a function of $\kappa_\parallel$, as

$$\lambda_\parallel = 3\kappa_\parallel/c \sim 3 \text{ pc} \gg R_L = 2 \times 10^{-3} \text{ pc}, \tag{12}$$

where $R_L$ is the Larmor radius of 5 TeV protons in a 3 μG magnetic field. This implies a very large Bohm factor, which is defined as the ratio of mean free path to Larmor radius, suggesting that the cross-field diffusion is negligibly smaller than the drift effect. In this case, the perpendicular UDF is produced solely from the drift flux, as

$$\eta_\perp \sim R_L/L', \tag{13}$$

giving an alternative scale length $L' \sim 2$ pc with $\eta_\perp \sim 10^{-3}$. To interpret the anisotropy seen by Tibet III, therefore, we need a structure with scale of $L' \sim 2$ pc, which is much smaller than that of $L \sim 3.2$ kpc for entire Galaxy. This is why paper I proposed a LISMF model based on a local structure associated with the local interstellar cloud (LIC) with a radius of ~3 pc [14].

In this paper, we analyzed the sidereal anisotropy of GCR intensity in the sub-TeV region observed with the THN of underground muon detectors. The THN can observe not only the northern hemisphere, but also the entire sky in the southern hemisphere which is not observed by the Tibet III experiment monitoring the multi-TeV GCR intensity. By applying a best-fit analysis based on the same model anisotropy in space applied for Tibet III data by paper I, we find that the best-fit orientation of reference axis of the anisotropy is quite similar for each experiment, regardless of the order of magnitude difference in GCR energy. This is apparent in Figure 4, which displays celestial sky-maps derived from best fitting the same model anisotropy to the THN data (upper panel) and Tibet III data (lower panel). This similarity implies, at least indirectly, that the conclusion derived by paper I based on the Tibet III data only in the northern hemisphere is not seriously biased. The best-fit amplitudes of the anisotropy, on the other hand, are only one third or less of those from Tibet III, indicating an attenuation possibly due to solar modulation [20]. The rigidity dependence of the anisotropy amplitude in the sub-TeV region is consistent with the spectrum ($P_m^{0.7}$) reported by Hall *et al.* [11], smoothly extending to the Tibet III result in the multi-TeV region. The amplitude at higher energies appears almost constant or gradually decreasing with the increasing rigidity.

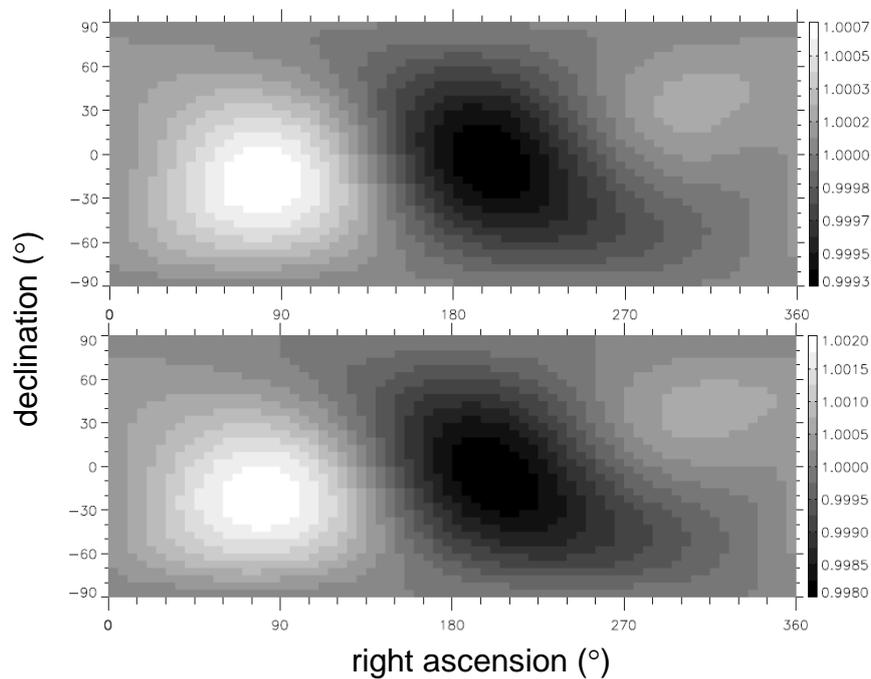

Fig. 4. Sky maps of the model anisotropy derived from best-fitting to sidereal daily variations observed with the THN (top) and the Tibet III experiment (bottom). Each panel displays the GCR intensity in 5°×5° pixels as a function of the right ascension on the horizontal axis and the declination on the vertical axis in a gray-scale format. The saturations in the top and bottom panels occur at ±0.0007 and ±0.002 respectively, as indicated by color code bars on the right side. Note that the average in each 5° declination band is subtracted from the intensity in each pixel.

The air shower experiment has a response not only to primary nuclei component in GCRs but also to high energy gamma rays, to which underground muon detectors have only negligible response. It is possible, therefore, that the anisotropy by Tibet III contains a significant contribution from primary gamma rays. The Super Kamiokande experiment, a deep underground detector capable for observing multi-TeV GCRs by measuring muons, however, has recently reported a large-scale anisotropy in agreement with the Tibet III result [21]. This implies that the large-scale anisotropy observed by Tibet III is not due to high energy gamma rays, but due to the anisotropy of the charged component of primary cosmic rays. This is also consistent with the attenuation of amplitude in the sub-TeV region deduced in this paper, as the attenuation due to solar modulation cannot be expected for an anisotropy due to primary gamma rays.

The model anisotropy with the best-fit parameters derived in this paper, on the other hand, does not fully reproduce the NS asymmetric feature of the observed anisotropy. The observed phase of intensity maximum shifts to earlier hours as the viewing latitude moves southward over the equator, as reported earlier from the THN and Tibet III experiments [10-11] [18]. A localized region of excess 10 TeV cosmic rays is also seen in Tibet III data at around the right ascension of ~70° and the declination of ~15° in agreement with a recent report by Milagro experiment [22]. These observed features need to be modeled further.

### Acknowledgments

The THN observations were supported by research grants from the Japanese Ministry of Education, Science, Sports and Culture and the Australian Research Council. The authors gratefully acknowledge the help received from the Tasmanian Hydro-Electric Commission for their invaluable assistance during periods of maintenance work at Liapootah. They also appreciate financial support from the Mitsutoyo Co. The observations at Matsushiro have been supported by Shinshu University, while the observations at Liapootah were maintained by the University of Tasmania and the Australian Antarctic Division.